\documentclass[12pt,preprint]{aastex}
\def\insertplot#1#2#3#4#5#6#7{
\vskip 10pt\nobreak\hbox to \hsize{\hss\dimen0=#3in\hbox to #6\dimen0{%
\dimen0=#2in\vbox to #6\dimen0{\vss
\special{ps: plotfile #1}
\special{ps::[end]
  PGPLOT restore
}
}\hss}\hss}\vskip 10pt}
\begin{document}
\title{Evidence for a Turnover in the Initial Mass Function of Low--Mass Stars and \\ Substellar Objects:  Analysis from an Ensemble of Young Clusters}
\author{M. Andersen}
\affil{ mortena@ipac.caltech.edu}

\and

\author{ M. R.  Meyer, J. Greissl. \& A. Aversa}
\affil{Steward Observatory, University of Arizona, Tucson, AZ 85721} 
\begin{abstract} 
We present a combined analysis of the low-mass initial mass function (IMF) for seven  star-forming regions. 
We first demonstrate that the ratios of stars to brown dwarfs are consistent with a single underlying IMF. 
By assuming that the underlying IMF is the same for all seven clusters and by combining the ratio of stars to brown dwarfs from each cluster we constrain the shape of the brown dwarf IMF and find it to be   consistent with a lognormal IMF . 
This provides the strongest constraint yet that the substellar  IMF turns over ($\frac{dN}{dM}\propto M^{-\alpha}$, $\alpha < 0$). 

\end{abstract} 


\section {Introduction}

Speculations concerning the existence and frequency of brown dwarfs
can be traced to before the introduction of the term \citep{kumar,hayashi}.  
Since then, 
wide-field surveys have   uncovered hundreds of candidates in the field and 
revealed two new spectral types, the L and T dwarfs \citep{kirkpatrick}. 
Yet the frequency of brown dwarfs compared to stars has remained a topic
of confusion and debate.  In a pioneering work, \citet{reid} 
attempted the first census of the substellar initial mass function (IMF) based on 
results from the Two Micron All Sky Survey \citep{skrutskie}.  
They presented evidence for a low-mass IMF that was more shallow than a Salpeter \citep{salpeter} slope, 
suggesting that brown dwarfs were not a significant contributor to dark matter. 
\citet{allen} used a Bayesian approach  to constrain the power-law 
slope below 0.08 $M_\odot$ to be in the 
range $-0.6 < \alpha < 0.6$ with a confidence level of 60\%, where a Salpeter slope is $\alpha=2.35$. 
These results indicate that, although brown dwarfs 
do not contribute significantly 
to the mass of typical stellar populations, 
they might still be as abundant as stars \citep{chabrier02}.  

The classical approach to deriving 
the mass function for stars and substellar objects is
to take an observed luminosity function and  apply a mass-luminosity
relationship in order to  derive the present-day mass function.  Then, 
corrections, based on the theory of stellar evolution, permit one
to estimate an {\it initial} mass function from the present-day mass function (see e.g. \citet{scalo,kroupa,chabrier} for 
complete descriptions of this process).  The confounding variable in 
these analyses is the star formation history of the Galactic disk, 
which is vital for substellar objects whose mass--luminosity
relationship evolves with  time.

A different  approach is to use star clusters of known age
as laboratories to measure the IMF\@.  
Open clusters 
 are  in principle good candidates because of their richness. 
Yet they suffer from the effects of dynamical evolution, 
mass segregation, and evaporation \citep[e.g.][]{ladalada}.
Young ($<$ 10 Myr) embedded clusters
 are attractive alternatives 
as they are compact and rich (from hundreds to thousands of stars within 
0.3--1 pc), and yet to emerge as unbound OB/T 
associations, 
 and the low mass objects are  10--1000 times more luminous 
than their   older open cluster  counterparts (0.1--16 Gyr) because they shrink  and cool as they age.  

Indeed, embedded clusters have been 
the targets of aggressive  photometric and spectroscopic
surveys in an attempt to search for variations in the IMF as
a function of initial conditions.  \citet{meyer00} found
that the ratio of high-mass (1--10 $M_{\odot}$) 
to low-mass (0.1--1 $M_{\odot}$)  stars for an ensemble of 
young clusters within 1 kpc was consistent with (1)  each 
other and (2)  having been drawn from the
field star IMF\@.  More recent studies have pushed well into 
the substellar mass regime (see \citet{luhman07a} for a recent
review). 
There have been some claims for variations in the brown dwarf IMF between nearby star-forming regions. 
 \citet{briceno} argued that the low-density 
Taurus dark cloud had a dearth of brown dwarfs compared to the 
rich Orion Nebula Cluster (ONC). 
However, this preliminary result has been updated as 
additional data have become available and as the statistics improved for both clusters \citep{guieu,slesnick}.

Here we use observations of seven nearby star clusters to constrain the combined brown dwarf IMF. 
In section 2, we describe the data, illustrate that there is no strong evidence for variation in the substellar IMF between the star-forming regions, and outline our approach to constrain the low-mass IMF.  
In section 3 we present our results, and in 
section 4 we discuss our results in the context of previous
work as well as theories of star (and substellar object)
formation. 

\section{The Approach} 

We have compiled the ratio of stars to brown dwarfs in nearby, well-studied young embedded clusters and the Pleiades. 
The regions included in this study are described briefly below, where the ratio of stars (0.08--1.0 $M_\odot$) to brown dwarfs (0.03--0.08 $M_\odot$) is calculated. 
For all the regions, we consider the {\it system} IMF, uncorrected for multiplicity within 200 AU. 
The sample is focused on  embedded clusters, in which  spectroscopy has been used to determine the age of the cluster, field star contamination has been taken into account, and an extinction-limited sample has been defined. 
Furthermore, we have included the Pleiades, because it is one of the best-studied open clusters and bacause its  substellar IMF has been estimated. 
The break point at 0.08 $M_\odot$ has been adopted in accordance with the break point for the \citet{kroupa} IMF, similar to the  characteristic mass in the  \citet{chabrier} single object IMF\@. 
Only a few of the clusters adopted here have the IMF derived in an extinction-limited sample reaching 0.02 $M_\odot$ and we have opted for 0.03 $M_\odot$ as a lower mass limit to obtain a larger  sample of clusters. 

{ \it Taurus.}
\citet{luhman04} imaged a 4 deg$^2$ region of Taurus that focused on the denser filaments, to identify cluster candidates. 
Candidates were confirmed as cluster members,by use of follow-up intermediate-resolutionoptical spectroscopy, on the basis of their effective temperature, luminosity, and spectral features. 
In total, 112 objects were confirmed members with  derived masses between 0.03 and 1.0 $M_\odot$ and  extinctions A$_\mathrm{V} \le 4$ mag. 
Some  96 objects were stars and 16 were brown dwarfs. 
Thus, the ratio of stars to brown dwarfs   in Taurus was found to be  $R=96/16=6.0^{+2.6}_{-2.0}$ where the errors are estimated using the method of \citet{gehrels86}. 

{\it IC 348.}
\citet{luhman03b} imaged a 42\arcmin$\times$28\arcmin\ region of the IC 348 cluster to identify cluster candidates. 
By the use of intermediate-resolution spectroscopy, most of the candidates were confirmed as cluster members, on the basis of  their effective temperature, luminosity, and spectral features, which indicated that the objects were  young. 
In total, \citet{luhman03b} found 168 cluster members with  masses between 0.03 and 1.0 $M_\odot$ and  extinctions A$_\mathrm{V} \le 4$ mag. 
The ratio of stars to brown dwarfs was found to be $R=8.3^{+3.3}_{-2.6}$. 

{\it Mon R2.}
\citet{andersen} imaged the central 1\arcmin$\times$1\arcmin\ of the embedded cluster associated with Mon R2 by utilizing the Near-Infrared Camera andd Multi-Object Spectrometer on board the {\it Hubble Space Telescope} ({\it HST}). 
An extinction-limited sample A$_\mathrm{V}\le 10$ mag was defined and  
a total of 19 objects were detected with masses between 0.03 and 1 $M_\odot$. 
The ratio of stars to brown dwarfs was found to be $R=8.5^{+13.6}_{-5.8}$. 

{\it Chameleon 1.}
 \citet{luhman07b} obtained an extinction-limited sample  in Chameleon 1 that was complete down to 0.01 $M_\odot$ for A$_\mathrm{V}\le 5$ mag, by use of   observations of a 0.22\arcdeg $\times$0.28\arcdeg\ region with the Advanced Camera for Surveyes on board {\it HST} and a subsequent spectroscopic follow-up of cluster member candidates. 
The sub-sample from 0.03 to $1$ $M_\odot$ includes  24 objects and the ratio $R$ was found to be $R=4.0^{+3.7}_{-2.1}$. 

{\it Pleiades. }
The Pleiades is one of the best-studied open clusters, and numerous derivations of the IMF have been published. 
Here we focus on the  survey by \citet{moraux} who covered a 6.4 deg$^2$ region of the Pleiades. 
The survey had a saturation limit of 0.48 $M_\odot$.  
For higher masses, the survey was combined   with a mass function built using the  \citet{prosser98} database.  
The Pleiades suffer relatively low ($A_\mathrm{V} < 1$ mag), mostly uniform, extinction, with negligible impact on 
the completeness of this sample, so we did not apply a reddening criterion. 
The ratio of stars to brown dwarfs was found to be $R=4.9^{+1.5}_{-1.2}$. 

{\it The Orion Nebular Cluster.}
The ONC 
 has been the subject of extensive
studies \citep{hillenbrand,hillenbrandcarpenter,luhman00,muench}.   
We take the adopted
ratio of stars to substellar objects from the 
study of \citet{slesnick}. 
The total sample, covering the central 5.1\arcmin$\times$5.1\arcmin,  
contains approximately 200 objects with masses
between 0.02 and 0.6 $M_{\odot}$ and A$_\mathrm{V} \le 15$ mag.  
Using their Figure 14, 
and extrapolating the slope from 0.08--0.6 
to 1.0 $M_{\odot}$ (one additional bin in their
plot), we arrive at a ratio of stars to substellar objects
of $R=3.3^{+0.8}_{-0.7}$.

{\it NGC 2024.}
The ratio of stars to brown dwarfs in NGC 2024 was found by \citet{levine} from their photometric and spectroscopic  study, covering the central 10\arcmin$\times$10\arcmin\@.
They assigned masses to the photometric objects  
 on the basis of the mass distribution in each
magnitude bin, determined from the spectroscopic sample 
as in \citet{slesnick}.  The result was that a total of 148 objects in their survey area has  masses
between 0.02 and 1 $M_{\odot}$ and extinctions A$_\mathrm{V}\le$ 15 mag. 
Based on their Figure 9, we find  that
there are 27 objects between 0.03 and 0.08 $M_{\odot}$
resulting in a ratio of stars to substellar objects
of $R= 3.8^{+2.1}_{-1.5}$. 

\begin{figure}
\epsscale{1.}
\plotone{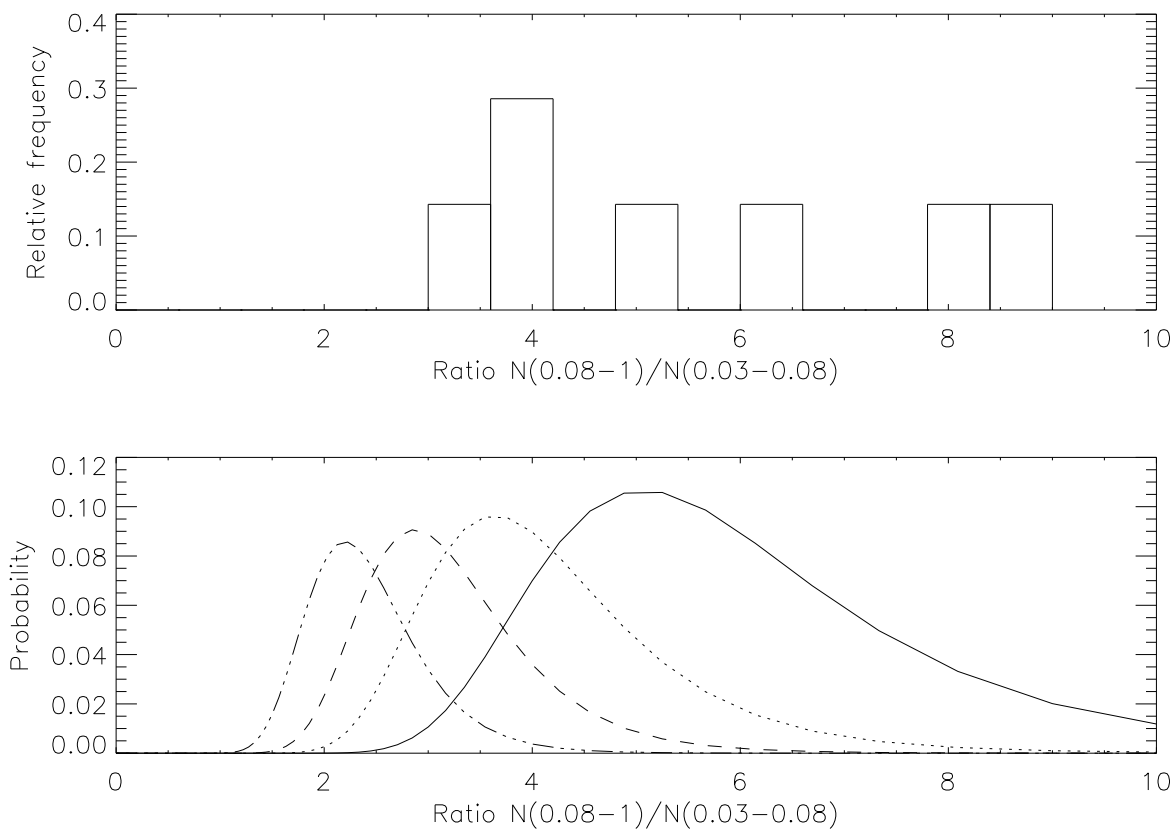}
\caption{{\it Top panel}: Histogram of the observed ratios of stars to brown dwarfs described in the text and summarized in Table~\ref{results}. 
{\it Bottom panel}: Binomial distribution for a cluster with 100 objects drawn from either the Chabrier ({\it solid line}), the falling ($\alpha=-0.6$, {\it dotted line}), the flat ($\alpha=0$; {\it long-dashed}), or the  rising ($\alpha=0.6$; {\it long-dash-dotted line}) IMF. 
Distributions that
continue to rise in linear mass units below the hydrogen 
burning limit are least consistent with the observations.}

\label{figure1}
\end{figure}

Table~\ref{results} shows the ratio of stars to brown dwarfs for nearby embedded clusters and the Pleiades, as 
described above, and the distribution of ratios is shown in Fig,~\ref{figure1}. 
The weighted mean of the ratios is found to be 4.3, and the standard deviation of the weighted mean   is 1.6. 
All of the measurements presented are consistent with the weighted mean within 2$\sigma$. 
There is thus  little evidence for variation in the low-mass IMF between the different regions and we have adopted the hypothesis that the IMF is  universal.
Under this assumption, the complete set of IMF determinations can be combined to place  constraints that are stronger than for each of the individual measurements.


\section{The Results}

For each cluster, we have  calculated the probability of obtaining the observed ratio of stars to brown dwarfs for a given IMF
or greater. 
The ratio  of stars to brown dwarfs drawn from a given sample size with an assumed  IMF is  determined by the binomial theorem. 
The predicted distribution of ratios from both segmented power-laws and a   \citet[][$\frac{dN}{d\log m}\propto \exp{\frac{(\log m-\log m_0)^2}{2\sigma^2}}$, $m_0=0.25$, $\sigma=0.55$]{chabrier05} lognormal  IMF for a cluster of 100 objects 
with unresolved binaries is shown in the lower  panel in Fig.~\ref{figure1}. 
The peak mass in the lognormal is slightly higher, and the width is slightly more narrow than is presented in \citet{chabrier}. 
The change in the best-fit parameters in \citet{chabrier05} is due to an updated luminosity function \citep{reid02}. 
A similar increase in the peak mass has  been suggested by \citep{covey}. 

The slope of the segmented power-law between  0.08 and 1.0 $M_\odot$ was chosen to be $\alpha=1.3$, and the slope has been varied below 0.08 $M_\odot$ in the range  $-0.6 < \alpha < 0.6$, which is the 60\% confidence interval presented by \citet{allen}.  
It is clear that the rising and flat IMFs ($\alpha=0.6$, and $0.0$, respectively) are difficult to reconcile with the observed distribution of ratios. 
We have quantitatively assessed the likelihood of obtaining the observed ratios from an assumed IMF as follows. 
For each of the seven measurements, the probability of obtaining that ratio or higher, assuming an underlying IMF, is calculated by adopting the binomial theorem. 
The product of the seven probabilities is then calculated. 
We find these values, which we refer to as the binomial tail product, or BTP, to be  $0.0012$, $2.2\times10^{-8}$, $1.8\times10^{-14}$, and  $1.0\times10^{-24}$, for a Chabrier, falling, flat, and  rising IMF, respectively. 
If each cluster sample was drawn from the assumed underlying IMF, and if  each cluster had an infinite number of objects, we would expect the combined product of this statistic for a sample of seven clusters to be $0.5^7=7.8\times10^{-3}$. 
The lognormal IMF appears to reproduce the observed ratios best, followed by the falling power-law IMF\@. 

\begin{figure}
\epsscale{1.}
\plotone{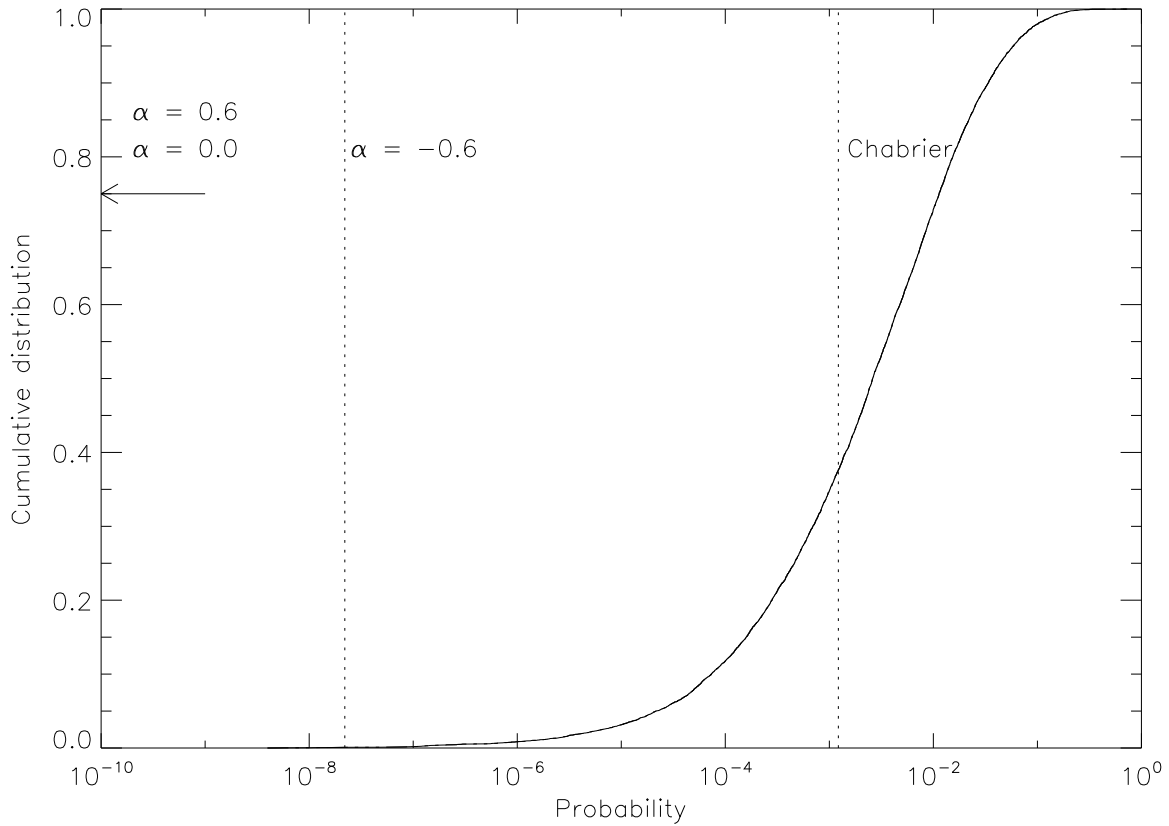}
\caption{Test of the distribution of the product of  probabilities if seven clusters are randomly drawn or higher from a Chabrier IMF\@. For each the probability of obtaining the observed ratio of stars to brown dwarfs is calculated and the product of the seven probabilities is determined for each of the 10\ 000 simulations. 
The vertical lines indicate the combined probability of obtaining the observed ratios of stars to brown dwarfs for the Chabrier IMF ({\it right vertical dotted} line) and the power-law IMF that is falling in linear units in the brown dwarf regime ($\alpha=0.6$; {\it left vertical dotted} line). 
The probabilities for the flat and rising IMF are both outside the plotted range and did not happen in any of the Monte Carlo simulations.}
\label{figure2}
\end{figure}

How consistent are the measured ratios with a Chabrier IMF and with what confidence can  other  IMFs be ruled out?
We have investigated that question by performing Monte Carlo simulations. 
We created an artificial set of seven clusters, each containing 100 objects (the median number of objects in our sample). 
The 100 objects are then  assigned masses according to the assumed underlying IMF, and the ratio of stars to brown dwarfs for each cluster is determined. 
For each of the ratios, the probability of observing that value or higher is calculated and the seven probabilities are multiplied, as was done for the observed set of clusters. 
The BTP for the observed clusters is then compared with the distribution of BTPs just derived. 
Because each factor in the BTP is drawn from a binomial distribution (of varying shapes), each 
IMF gives the same expected distribution of BTPs. 
Figure~\ref{figure2} shows the cumulative distribution of BTPs for a set of 10,000 simulations. 

Overplotted are the probabilities obtained above for the observed set of clusters assuming the four different underlying IMFs. 
We find that  37\% of the simulations have a probability equal to or lower than what was found assuming a Chabrier IMF, and in only $\sim$0.05\%-0.1\% of the simulations is the probability equal to or lower than found assuming a falling power-law IMF\@. 
In none of the simulations did the low probabilities for the flat or rising power-law IMFs occur 
(P $<$ 0.01 \%). 
The results indicate that the IMF is falling in the brown dwarf regime and that the Chabrier IMF is  consistent with the observations. 
\section{Discussion}

The results on the IMF presented here are based on the system IMF, including binaries unresolved  within 200 AU. 
As such, they may be difficult to compare directly with the locally derived (within 20 pc) field IMF 
discussed in \citet{allen} that suffers from a much smaller fraction of unresolved binaries. 
Yet the overall binary  frequency for ultra-cool dwarfs (M6 and later) appears to be low \citep[$\sim$20\%,][]{burgasser}, 
and furthermore the {\it relative} number of companions with separations $>$15 AU and mass ratios 
q $>$ 0.4 may be extremely low around very cool stars $\sim$ 1\%; \citet{allen_new}. 

Indeed, if the companion mass ratio distribution follows the Chabrier IMF at wide separations, then one 
could expect fewer very low mass companions as one surveys progressively lower mass primaries \citep[e.g.][]{siegler},  
consistent with the observations by \citet{mccarthy}. 
If the  IMF follows a Chabrier IMF in  the brown dwarf regime below 0.03 $M_\odot$ 
(say, down to the opacity limit for fragmentation of $\sim$ 0.001-0.004 $M_\odot$; \citet{whitworth}, 
then the number of stars below 1 $M_\odot$ will outnumber brown dwarfs 4.7 to 1. 

The sense of our results, that the mass function is falling in the BD regime, is consistent with various ideas put forward to 
explain the shape of the IMF (\citet{bonnell07} and references therein).  Building on the ideas of \citet{larson05}, 
\citet{bonnellbate} produced an IMF that is only weakly dependent on the Jeans mass  through dynamical interactions in the cluster. 
However, \citet{allen_new} show that the turbulent fragmentation models by  \citet{batebonnell} predict too few low-mass binary systems. 
\citet{goodwin}, on the other hand, suggest that the IMF should peak at higher masses in regions with low turbulence  
(e.g. Taurus) which would result in a higher ratio of stars to brown dwarfs. 
The lack of a strong variations in the ratio of stars to brown dwarfs is a problem for the turbulence models in general; 
for example, magnetic turbulence models predict strong variations in the low-mass IMF as a function of Mach number and 
density \citep{padoan}. 
If the preliminary results indicated here are borne out through further
observations, then models that depend only weakly on initial conditions would be required (e.g. \citet{adams96}; \citet{hennebelle}). 

Possible IMF variations at least within 1 kpc are smaller than 
can be detected by comparing the currently observed  clusters. 
Thus, there are two challenges in detecting IMF variations: (1) 
One needs clusters with a well-sampled population to minimize the inherently stochastic nature of  populating an IMF, and  (2)  a larger set of clusters is needed to detect even small IMF variations with initial conditions. 
Although it appears that the variations in the IMF down to 30 $M_\mathrm{Jup}$ are modest, we still expect that variations will be seen at the lowest masses  where the opacity limit for fragmentation can be reached \citep{lowlynden-bell} and the metallicity of the star forming region could be imprinted in the lower mass limit. 

\acknowledgements{We thank Joanna Levine, Kevin Luhman, and Cathy Slesnick for helpful discussions, 
as well as Neill Reid, Charles Lada, and Pavel Kroupa for comments on a draft of this Letter. 
The referee is acknowledged for a very fast response and for suggestions that improved the manuscript. 
Finally, we thank the organizers of the Cool Stars 14  Splinter Session entitled 
{The Formation of Low-Mass Protostars and Proto-Brown Dwarfs} 
for the opportunity to present a  preliminary version of this work. 
MRM gratefully acknowledge the support of a Cottrell Scholar
award from the Research Corporation, NASA grant 
GO-9846 from the Space Telescope Science Institute, 
and the Arizona Space Grant Consortium. }

\clearpage

\begin{deluxetable}{ccccccllll}
\rotate
\tablecolumns{10}
\tablewidth{0pc}
\tablecaption{Ratio of Stars to substellar Objects in Young Clusters. 
The distance, age, number of objects in the sample, and the extinction limit used for the embedded clusters are given. 
The four last columns gives the probability of the observed ratio having been drawn from the assumed IMFs.}

\centering
\tablehead{
\multicolumn{1}{c}{Cluster}   &
\multicolumn{1}{c}{Dist.} &
\multicolumn{1}{c}{ Age } &
\multicolumn{1}{c}{$N_{obj}$}&
\multicolumn{1}{c}{Max $A_{V}$}&
\multicolumn{1}{c}{$R=\frac{N(0.08-1.0)}{N(0.03-0.08)}$}  & 
\multicolumn{1}{l}{P($R\ge R_{obs}$) } & 
\multicolumn{1}{l}{P($R\ge R_{obs}$) } & 
\multicolumn{1}{l}{P($R\ge R_{obs}$) } & 
\multicolumn{1}{l}{P($R\ge R_{obs}$) }\\ 
\multicolumn{1}{c}{}   &
\multicolumn{1}{c}{(pc) } &
\multicolumn{1}{c}{ (Myr) } &
\multicolumn{1}{c}{}&
\multicolumn{1}{c}{Mag}&
\multicolumn{1}{}{}  &
\multicolumn{1}{l}{Chabrier} & 
\multicolumn{1}{l}{$\alpha=-0.6$} & 
\multicolumn{1}{l}{$\alpha=0$} & 
\multicolumn{1}{l}{$\alpha=0.6$}}
\startdata

Taurus & 140 & 1--3  & 112 & 4.0 & $6.0^{+2.6}_{-2.0}$ & 0.286 & 0.030 & 0.002 & 2.47$\cdot10^{-5}$  \\
ONC & 480 & 1  & 185& 2.0 & $3.3^{+0.8}_{-0.7}$ & 0.907 & 0.744 &  0.365 & 0.066 \\
Mon R2 & 830 & 1  & 19  & 10 & $8.5^{+13.6}_{-5.8}$  & 0.359 & 0.182 & 0.093 & 0.035 \\   
Chamaeleon & 160 & 2 & 24 & 5.0 & $4.0^{+3.7}_{-2.1}$ & 0.795 & 0.569 & 0.375 & 0.187  \\
Pleiades & 125 & 120 & 200 & 1.0 & $4.9^{+1.5}_{-1.2}$  & 0.560 & 0.056 & 0.002 & 7.39$\cdot10^{-6}$ \\   
NGC 2024 & 460 & 1 & 50 & 11.0 & $3.8^{+2.1}_{-1.5}$ & 0.877 & 0.591 & 0.317 & 0.097  \\
IC 348 & 315 & 2  & 168 & 4.0 & $8.3^{+3.3}_{-2.6}$  & 0.031 & 3.00$\cdot10^{-4}$ & 1.88$\cdot10^{-6}$ & 8.21$\cdot10^{-10}$ \\

\enddata
\label{results}

\end{deluxetable}

\clearpage

\end{document}